
%
%
\def\rhonaught{ \rho_{0} }
\def\phiR{ \,{}^{R}\!\!\phi }
\def\phiI{ \,{}^{I}\!\!\phi }
\def\phiH{ \phi_{H} }
\def\phiG{ \phi_{G} }
\def\phiHmin{ \phi_{H min} }
\def\phione{ \,{}^{1}\!\!\phi }
\def\phitwo{ \,{}^{2}\!\!\phi }
\def\fone{ \,{}^{1}\!\!f }
\def\ftwo{ \,{}^{2}\!\!f }
\def\fR{ \,{}^{R}\!\!f }
\def\fI{ \,{}^{I}\!\!f }
\def\frho{ \,{}^{\rho}\!\!f }

\def\monesq{ m_{1}^{2} }
\def\mtwosq{ m_{2}^{2} }
\def\mRsq{ m_{R}^{2} }
\def\moneeffsq{ m_{1\hbox{\the \scriptscriptfont0 eff}}^{2} }
\def\mtwoeffsq{ m_{2\hbox{\the \scriptscriptfont0 eff}}^{2} }
\def\mReffsq{ m_{R\hbox{\the \scriptscriptfont0 eff}}^{2} }
\def\mHeffsq{ m_{H\hbox{\the \scriptscriptfont0 eff}}^{2} }
\def\mrhoeffsq{ m_{\rho\hbox{\the \scriptscriptfont0 eff}}^{2} }
\def\oneDelta#1{ \,{}^{1}\!\!\Delta (#1) }
\def\twoDelta#1{ \,{}^{2}\!\!\Delta (#1) }
\def\Pitotvac#1{%
\Pi_{\vbox{\offinterlineskip
\hbox{\the\scriptscriptfont0 TOTAL}%
\vskip 1 true pt%
\hbox{\the\scriptscriptfont0 vacuum}%
}}^{( #1 )} (P^2)}
\def\Pitottherm#1{%
\Pi_{\vbox{\offinterlineskip
\hbox{\the\scriptscriptfont0 TOTAL}%
\vskip 1 true pt%
\hbox{\the\scriptscriptfont0 thermal}%
}}^{( #1 )} (P^2)}
\def\atpzero{\Biggm\vert_{P = ( 0 , \vec{0} )}}
\def\atthermal{\Biggm\vert_{\hbox{\the\scriptscriptfont0 thermal}}}
\def\onshell{\Biggm\vert_{\hbox{\the\scriptscriptfont0 on-shell}}}
\def\atthermalatrest{\Biggm\vert_{\vbox{\offinterlineskip
\hbox{\the\scriptscriptfont0 thermal}%
\vskip 1 true pt%
\hbox{\the\scriptscriptfont0 at rest}%
}}}
\def\atresthightemp{\Biggm\vert_{\vbox{\offinterlineskip
\hbox{\the\scriptscriptfont0 at rest}%
\vskip 1 true pt%
\hbox{\the\scriptscriptfont0 high temp.}%
}}}
\def\atresthightemponshell{\Biggm\vert_{\vbox{\offinterlineskip
\hbox{\the\scriptscriptfont0 at rest}%
\vskip 1 true pt%
\hbox{\the\scriptscriptfont0 high temp.}%
\vskip 1 true pt%
\hbox{\the\scriptscriptfont0 on-shell}%
}}}
\def\onpertshell{\Biggm\vert_{\vbox{\offinterlineskip
\hbox{$\scriptstyle p^2 = m^2$}%
\vskip 1 true pt%
\hbox{$\scriptstyle \rho^2 = \overline{\rho}_{(0)}^2$}%
}}}
\def\Lag{{\cal L}}
\def\Lagcl{{\cal L}_{\hbox{\the\scriptscriptfont0 constant + linear}}}
\def\Lagqd{{\cal L}_{\hbox{\the\scriptscriptfont0 quadratic}}}
\def\Lagcu{{\cal L}_{\hbox{\the\scriptscriptfont0 cubic}}}
\def\Lagqr{{\cal L}_{\hbox{\the\scriptscriptfont0 quartic}}}
\def\Lagho{{\cal L}_{\hbox{\the\scriptscriptfont0 higher-order}}}
\def\Laggh{{\cal L}_{\hbox{\the\scriptscriptfont0 ghost}}}
\def\Laggf{{\cal L}_{\vbox{\offinterlineskip
\hbox{\the\scriptscriptfont0 GAUGE}%
\vskip 1 true pt%
\hbox{\the\scriptscriptfont0 FIXING}%
}}}
\def\pbar{\overline{p}}
\def\cbar{ \bar{c} }
\def\sumintk{ {{1}\over{\beta}} {{\sum}\atop{n_k}}
               \int {{d \vec{k}}\over{ (2 \pi)^3 }} }
%
%
%
\def\Higgsaone{%
\vbox to 40 pt{\vss
\hbox to 70 pt{\hss
\hss}}}
\def\Higgsatwo{%
\vbox to 40 pt{\vss
\hbox to 70 pt{\hss
\hss}}}
\def\Higgsbone{%
\vbox to 40 pt{\vss
\hbox to 70 pt{\hss
\hss}}}
\def\Higgsbtwo{%
\vbox to 40 pt{\vss
\hbox to 70 pt{\hss
\hss}}}
\def\Higgsbg{%
\vbox to 40 pt{\vss
\hbox to 70 pt{\hss
\hss}}}
\def\Higgscone{%
\vbox to 40 pt{\vss
\hbox to 70 pt{\hss
\hss}\vss}}
\def\Higgsctwo{%
\vbox to 40 pt{\vss
\hbox to 70 pt{\hss
\hss}\vss}}
\def\Higgscg{%
\vbox to 40 pt{\vss
\hbox to 70 pt{\hss
\hss}\vss}}
\def\Goldaone{%
\vbox to 40 pt{\vss
\hbox to 70 pt{\hss
\hss}}}
\def\Goldatwo{%
\vbox to 40 pt{\vss
\hbox to 70 pt{\hss
\hss}}}
\def\Goldag{%
\vbox to 40 pt{\vss
\hbox to 70 pt{\hss
\hss}}}
\def\Goldbone{%
\vbox to 40 pt{\vss
\hbox to 70 pt{\hss
\hss}}}
\def\Goldbtwo{%
\vbox to 40 pt{\vss
\hbox to 70 pt{\hss
\hss}}}
\def\Goldbg{%
\vbox to 40 pt{\vss
\hbox to 70 pt{\hss
\hss}}}
\def\Goldconetwo{%
\vbox to 40 pt{\vss
\hbox to 70 pt{\hss
\hss}\vss}}
\def\Higgsprop{%
\vbox to 10 pt{\vss
\hbox to 50 pt{\hss
\hss}\vss}}
\def\Goldprop{%
\vbox to 10 pt{\vss
\hbox to 50 pt{\hss
\hss}\vss}}
\def\Ghostprop{%
\vbox to 10 pt{\vss
\hbox to 50 pt{\hss
\hss}\vss}}
\def\gone{%
\vbox to 30 pt{\vss
\hbox to 50 pt{\hss
\hss}\vss}}
\def\gonetwo{%
\vbox to 30 pt{\vss
\hbox to 50 pt{\hss
\hss}\vss}}
\def\gonecbarc{%
\vbox to 30 pt{\vss
\hbox to 50 pt{\hss
\hss}\vss}}
\def\lambdaone{%
\vbox to 30 pt{\vss
\hbox to 50 pt{\hss
\hss}\vss}}
\def\lambdaonetwo{%
\vbox to 30 pt{\vss
\hbox to 50 pt{\hss
\hss}\vss}}
\def\lambdatwo{%
\vbox to 30 pt{\vss
\hbox to 50 pt{\hss
\hss}\vss}}
\def\lambdatwocbarc{%
\vbox to 30 pt{\vss
\hbox to 50 pt{\hss
\hss}\vss}}
\def\figurethree{
$$
\matrix{ {\rm a1)} \Higgsaone
& {\rm a2)} \Higgsatwo
& { }
\cr
{\rm b1)} \Higgsbone
& {\rm b2)} \Higgsbtwo
& {\rm bg)} \Higgsbg
\cr
{\rm c1)} \Higgscone
& {\rm c2)} \Higgsctwo
& {\rm cg)} \Higgscg
\cr}
$$
}
\def\figurefour{
$$
\matrix{ {\rm a1)} \Goldaone
& {\rm a2)} \Goldatwo
& {\rm ag)} \Goldag
\cr
{\rm b1)} \Goldbone
& {\rm b2)} \Goldbtwo
& {\rm bg)} \Goldbg
\cr
{\rm c12)} \Goldconetwo
& { }
& { }
\cr}
$$
}
%
\magnification 1200
\baselineskip 15pt
\hyphenation{re-norm-a-liza-bi-lity}
\centerline{\bf PARAMETERIZATION INVARIANCE AND THE RESOLUTION}
\centerline{\bf OF THE UNITARY GAUGE PUZZLE}
\bigskip\medskip
\centerline{P.F. Kelly}
\centerline{Center for Theoretical Physics, Laboratory for Nuclear
Science}
\centerline{Massachusetts Institute of Technology}
\centerline{Cambridge,  MA\  02139\ USA}
\medskip
\centerline{R. Kobes}
\centerline{Lab. de Phys. Th\'eor. ENSLAPP, BP 110}
\centerline{F-74941 Annecy-le-Vieux\ FRANCE}
\centerline{{\it and}}
\centerline{Winnipeg Institute for Theoretical Physics and Physics
Department}
\centerline{University of Winnipeg}
\centerline{Winnipeg,  MB\  R3B 2E9\ CANADA}
\medskip
\centerline{ G. Kunstatter}
\centerline{Winnipeg Institute for Theoretical Physics and Physics
Department}
\centerline{University of Winnipeg}
\centerline{Winnipeg,  MB\  R3B 2E9\ CANADA}
\bigskip
\noindent {\bf Abstract:}  We examine the calculation of the
critical temperature $T_c$ for the restoration
of a spontaneously broken symmetry.
Motivated by a set of recently developed gauge dependence
identities, we give a gauge and parameterization independent
definition of this temperature in terms of the physical mass
of the scalar particle as determined by the propagator pole.
As an explicit illustration, we consider the Abelian
Higgs model in the unitary gauge, where the usual definition
of the critical temperature based upon the effective potential
leads to an erroneous result.
We show how the gauge invariant definition reproduces the correct
result as found in the ``renormalizable'' parameterizations.
\bigskip
\line{\hfil MIT -- CTP -- 2299}
\line{\hfil WIN -- 93 -- 09}
\line{\hfil ENSLAPP -- A -- 470/94}
\line{\hfil NSF -- ITP --  94 -- 62}
\vfill\eject
\line{\bf 1. INTRODUCTION   \hfil }
\medskip
\par
That there are models in which symmetries, spontaneously
broken at low (zero) temperature, are restored at high
(finite) temperature was conjectured~[1], and explicitly
confirmed~[2,3] a number of years ago.
Quantitative investigations of the Abelian Higgs model
have exposed what
has become known as the {\sl Unitary Gauge Puzzle}~[2-6].
This puzzle arose when attempts were made to estimate the
transition or critical temperature for the restoration of the
symmetry; in the unitary gauge, a different
value for the critical temperature was obtained as compared
to those obtained in certain other gauges.
This difference has, in the past, been attributed to
pathologies of the unitary gauge and the related question of the
validity of the loop expansion.
The purpose of this paper is to
provide a detailed explanation for the discrepancy as well as a
proof that identical results for the critical temperature can in
principle be obtained in the unitary gauge and other gauges
provided that
one uses the correct definition of the critical temperature
and that it is evaluated in a self-consistent expansion.
\par
In order to set the stage for the detailed
presentation of this puzzle in the next section, we first
briefly review the usual approach to calculating the
critical temperature using the effective potential~[2].
This function, being defined for constant field
configurations, enables the survey of a large class of
translationally invariant expectation values of the field.
The ground state of the system is identified with
the global minimum of the effective potential, and
spontaneous symmetry breaking is signalled by a non-zero
vacuum expectation value (VEV) of the field.
Under the assumption that the transition to the symmetric phase
takes place in a continuous ({\it i.e.,} second-order) fashion,
the relevant VEVs must go to zero smoothly as the temperature is
raised to its critical value.
Thus, the ``origin'' must be a local maximum of the effective
potential in the broken phase, and
the condition signaling the onset of the
second-order phase transition is that the origin be both an
extremum and an inflection point of the effective potential.
\par
For concreteness, we consider a
complex scalar field parameterized in terms of a real
``Higgs'' mode $\phiH$ and a real ``Goldstone'' boson $\phiG$.
The order-parameter, whose vanishing signifies the onset of the
phase transition, is provisionally identified with the
effective thermal mass--squared of the Higgs field:
$$
\mHeffsq =
{{\partial^2 V( \phiH , \phiG , \beta )}\over{\partial^2 \phiH}}
\Bigm\vert_{ {\phiH = \phiHmin}\atop{\phiG = 0} }
\ .
\eqno{(1.1)}
$$
The second derivative is evaluated at the value of the VEVs which
gives rise to the minimum of the effective potential.
Thus, as the temperature
increases from below the critical temperature, both
$\mHeffsq$ and the VEV decrease towards zero.
Above the critical temperature, $\mHeffsq$ is positive and the VEV
is zero.
The critical temperature is thus defined to be that temperature
which causes $\mHeffsq$ to vanish at zero VEV:
$$
0 =
{{\partial^2 V( \phiH , \phiG , \beta_c )}\over{\partial^2 \phiH}}
\Bigm\vert_{ {\phiH = 0}\atop{\phiG = 0} }
\ .
\eqno{(1.2)}
$$
\par
This definition of the critical temperature, however, is
problematic for gauge theories, since the
effective potential, calculated from off-shell
Green functions at zero external four-momentum, is
gauge dependent~[2,7] and parameterization dependent~[8].
Several approaches may be taken to deal with these
dependences.
One is to consider only objects which are
manifestly gauge independent (in this context, the
Vilkovisky-DeWitt effective action~[9] is particularly suitable),
as the critical temperature defined by such means
must be gauge and parameterization invariant~[10].
Alternatively, one can look for signals for symmetry breaking
in terms of expectation values of gauge invariant composite
operators~[11].
Another approach is to try to extract gauge
invariant information
from gauge dependent objects like the usual effective
potential~[7]. In this regard the ``Ward-like'' Nielsen
identities are useful in determining the gauge and
parameterization dependence of quantities.
For example, the Nielson identities have been exploited
to show that symmetry breaking and restoration
are gauge invariant phenomena~[12].
Recently, they
have been generalized to obtain a set of algebraic
identities which describe the gauge dependence of all
$n$-point
functions derivable from the effective action~[13].
As we shall demonstrate, these identities guarantee
that the same result for the critical temperature
in any gauge or parameterization can be obtained
order-by-order in a self-consistent expansion.
\par
This paper is organized as follows. In Section 2, the
unitary gauge puzzle shall be briefly
reviewed and previous analyses of it discussed.
We shall emphasize results found by Ueda~[4] and
by Arnold, Braaten and Vokos (ABV)~[6].
Section 3 discusses the gauge dependence identities
and how they provide a formal resolution to the unitary
gauge puzzle to all loop-orders.
A two-loop calculation by ABV is then shown to corroborate
these formal arguments.
Further support for this resolution is presented in
Section 4, where we consider a complex $\Phi^4$ theory
which also suffers from a ``unitary gauge puzzle''
for different parameterizations of the complex field~[8].
Using the definition of the critical temperature discussed
in Section 3, we demonstrate that the correct result is
obtained in a large class of parameterizations.
This class includes a parameterization corresponding to
the one which gave rise to the original unitary gauge puzzle.
Section 5 contains a brief restatement of our conclusions.
\medskip
\goodbreak
\line{\bf 2.  THE UNITARY GAUGE PUZZLE  \hfil }
\par
Consider a complex scalar field $\Phi$ with a
renormalizable self-interaction,
$m^2 \vert\Phi\vert^2 + ( \lambda / 6 ) \vert\Phi\vert^4$,
respecting the inherent $U(1)$ global symmetry of the system.
By gauging this symmetry, a model of scalar electrodynamics is
obtained.
The model with $m^2 = - \mu^2$ is the
standard Abelian Higgs model with spontaneous symmetry breaking:
$$
\Lag  = ( {\cal D}_\mu \Phi )^{\dagger} ( {\cal D}^\mu \Phi )
+ \mu^2 \vert \Phi \vert^2
- {{\lambda}\over{3!}} ( \vert \Phi \vert^2 )^2
- {1\over{4}} F_{\mu\nu} F^{\mu\nu}
+ \Laggf
\ .
\eqno{(2.1)}
$$
The gauge-covariant derivative couples the
gauge field to the scalar with strength $e$:
$ {\cal D}_\mu \Phi = [ \partial_\mu + i e A_\mu ] \Phi $,
and the $U(1)$ field strength tensor
is $ F_{\mu\nu} = \partial_\mu A_\nu - \partial_\nu A_\mu $.
The gauge fixing term has been left unspecified at this point.
Before one is able to progress further, one must
choose a parameterization for the complex field.
There are two distinct options which
shall be discussed in turn.
\par
\smallskip
\goodbreak
\line{\bf 2.1) Cartesian parameterization \hfil }
\par
The complex field $\Phi$ may be decomposed into its real and
imaginary parts:
$$
\Phi = {{1}\over{\sqrt{2}}} \big[ \phiR + i \phiI \big]  \ .
\eqno{(2.2)}
$$
Spontaneous symmetry breaking produces an asymmetric vacuum
state about which the true quantum fields $\{ \fR , \fI \}$
fluctuate.
Without loss of generality, the entire zero temperature VEV
of the scalar field ({\it i.e.,} $\rhonaught / \sqrt{2}$) may be
assigned
to the real component:
$$
\phiR = \rhonaught + \fR  \ , \quad  \phiI = 0 + \fI \ .
\eqno{(2.3)}
$$
The vector field $A_\mu$ fluctuates about a zero VEV.
Note that the functional Jacobian corresponding to the change of
field variables from $\{ \Phi , \Phi^{*} \} \to \{ \fR , \fI \}$
is unity and thus the path integral measure is trivial.
\par
Feynman rules based upon the
Lagrangian written in terms of these shifted fields
may be derived in a straightforward manner.
In order to do this, a gauge fixing term must be specified,
but the results are independent of this choice.
The order-parameter as defined by the curvature of the
effective potential may also be expressed as the self-energy
of the Higgs field $\fR$ evaluated at zero external
four-momentum.
At one-loop:
$$
\mReffsq = \mRsq + \Pitotvac{RR} \atpzero
+ \Pitottherm{RR} \atpzero
\ .
\eqno{(2.4)}
$$
The $\mRsq$ term is the bare tree-level mass--squared
which is combined with the divergent one-loop
vacuum self-energy and renormalized in the usual way.
Explicit calculation of the finite temperature contribution
leads to, at leading-order, and in the high temperature limit,
$$
\mReffsq = 2 \mu^2 - {{\lambda}\over{9}} {1\over{\beta^2}}
- {1\over2} e^2 {1\over{\beta^2}}
+ O ( \beta^{-1} )
\ .
\eqno{(2.5)}
$$
The terms which arise from the scalar sector have coupling strength
$\lambda$, while those from the gauge-field loops couple with
$e^2$.
It is assumed that $\lambda$ and $e^2$ are of the
same order.
Setting the expression for the order-parameter $(2.5)$ to zero
results in an equation which leads to the
(correct) estimate of the critical temperature~[2]:
$$
0 = 2 \mu^2
- \bigg[ {{\lambda}\over{9}} + {1\over2} e^2 \bigg]
{1\over{\beta_c^2}}
+ O ( \beta_c^{-1} )
\ .
\eqno{(2.6)}
$$
\smallskip
\goodbreak
\line{\bf 2.2) Polar parameterization \hfil}
\par
An alternative approach to this analysis is to use
the magnitude and phase representation of the
complex field:
$$
\Phi = {1\over{\sqrt{2}}} \, \rho \, e^{i \theta / \rhonaught}
\ ,
\eqno{(2.7)}
$$
where $\langle \Phi \rangle = \rhonaught / \sqrt{2}$
is the VEV of the scalar field at zero temperature.
Furthermore, all explicit dependence upon the Goldstone mode $\theta$
is eliminated by choosing the ``unitary'' gauge:
$$
B_\mu = A_\mu + {1\over{g \rhonaught}} \partial_\mu \theta
\ .
\eqno{(2.8)}
$$
Spontaneous symmetry breaking affects the modulus field alone:
$$
\rho = \rhonaught + \frho
\ .
\eqno{(2.9)}
$$
Thus, only the ``physical'' degrees of freedom are present:
a massive scalar Higgs boson $\frho$,
and a massive vector meson $B_\mu$.
\par
Feynman rules for the Lagrangian expressed in terms of these
fields may then be derived.
Note that the gauge for the
vector field has been specified by $(2.8)$.
The model in this form has lost its property of being manifestly
renormalizable due to the large momentum behaviour of the
vector meson propagator.
The transformation of field variables gives rise to a
non-trivial functional Jacobian in the path--integral measure,
which may be incorporated into the Lagrangean in the
usual manner,
$$
\Laggh =
- \cbar \bigg( 1 + {{\frho}\over{\rhonaught}} \bigg) c
\ ,
\eqno{(2.10)}
$$
where $\{ \cbar , c \}$ are scalar fields obeying Fermi-Dirac
statistics.
Notwithstanding these complications, the calculation of the
one-loop self-energy of the Higgs field can be performed
in a straightforward manner.
At zero external four-momentum, one finds in
the high temperature limit the following
equation for the order parameter:
$$
\mrhoeffsq = 2 \mu^2
- {{\lambda}\over{12}} {1\over{\beta^2}}
- {1\over2}e^2 {1\over{\beta^2}}
+ O ( \beta^{-1} )
\ .
\eqno{(2.11)}
$$
Setting this expression equal to zero
produces an equation for the critical temperature,
$$
0 = 2 \mu^2
- \bigg[ {{\lambda}\over{12}} + {1\over2} e^2 \bigg]
{1\over{\beta_c^2}}
+ O ( \beta_c^{-1} )
\ .
\eqno{(2.12)}
$$
which does not agree with the equation $(2.6)$ obtained in the
Cartesian parameterization.
This disagreement constitutes the unitary gauge puzzle.
It is noteworthy that the contribution from the gauge
field sector is the same in both instances.
This suggests that the unitary gauge puzzle arises from
the scalar sector of the Abelian Higgs model, rather than
the gauge field sector.
\smallskip
\goodbreak
\line{\bf 2.3) Previous Analyses of the Unitary Gauge Puzzle \hfil}
\par
It has been suggested that the apparent non-renormalizability
of the Abelian Higgs model in the unitary gauge and the
concommitant question of the validity of the loop-expansion
is to blame for the unitary gauge puzzle.
However, Ueda~[4] observed that by using an improved
order-parameter [the second-derivative of the effective potential
plus ``$\Pi ( P^2 = - m^2 ) - \Pi ( P^2 = 0 )$'']
it was possible to obtain the same result for the transition
temperature in the unitary gauge as in other gauges.
As we shall see in the next section,
this corresponds to the pole--definition of the Higgs mass.
In the Cartesian parameterization, the one-loop self-energy
is independent of the external four-momentum and thus this
additional term vanishes.
However, in the unitary gauge the momentum-dependent vertices
give rise to a non-zero contribution
which exactly accounts for the difference between $(2.6)$ and
$(2.12)$.
\par
More recently, ABV re-analysed the unitary gauge
puzzle and determined two general
criteria which must be met for a calculation of the
critical temperature to yield a result which is meaningful.
The first criterion is that the order-parameter itself must be the
physical, gauge-invariant mass--squared of the Higgs field as
defined by the location of the propagator pole.
The second criterion is that the perturbative expansion and the
ultimate evaluation of physical quantities under investigation
must be self-consistently undertaken.
They performed a calculation of the critical temperature
in the unitary gauge in powers of $T / T_c$ valid for
low temperatures, by consistently extracting the leading
temperature dependence in a loop-expansion.
In this way, they argued that it was possible to
obtain agreement between the Cartesian gauge and the
unitary gauge values of the critical temperature,
provided that certain unsuppressed terms of the form
$(T/T_c)^{2n}$ for $n \ge 2$ which arose in the unitary
gauge calculation appeared with vanishing coefficients.
Through careful analysis they were able to show explicitly that the
$O( T^4 )$ terms did indeed cancel among
themselves up to two-loop-order. They were, however, unable to
prove or verify that this property persisted for all higher-orders.
Moreover, in the ABV method of analysis, it seemed that
cancellations were required between one- and two-loop
contributions to the self-energy.
\par
In the next section, a general framework justifying Ueda's approach
shall be presented.
Furthermore, in this context, the ``miraculous'' cancellations
found by ABV demonstrate the consistency
of the loop-expansion in the unitary gauge to two-loop-order.
\medskip
\goodbreak
\line{\bf 3.  GAUGE DEPENDENCE IDENTITIES  \hfil }
\par
The off-shell effective action and the effective potential
depend explicitly on the gauge and the field
parameterization in which they are calculated.
However, this dependence is controlled by a set of generalized
Ward identities that can be derived from the definition
of the effective action, or its Legendre transform, the generating
functional for connected diagrams.
A complete discussion of these identities,
and further references, can be found in~[13].
The generalized Ward identities yield an
equation that determines the gauge dependence of the two-point
function $D^{-1}_{ij}$ in a field theoretic model.
In the absence of sources, the equation reads:
$$
\Delta D^{-1}_{ij} = D^{-1}_{ik} \Delta X^k_j + D^{-1}_{jk}\Delta
X^k_i \ ,
\eqno(3.1)
$$
where we employ a condensed notation in which
$\{i,j,k, \ldots \}$ refer to all discrete field indices as
well as to continuous spacetime coordinates.
In the above, $\Delta X^i_j$ is a quantity that
can be calculated in perturbation theory.
At this stage, however, (3.1) is exact as it stands.
The crucial feature to note is that the change in the two-point
function is proportional to the inverse-propagator itself.
\par
For the scalar mode in the spontaneously broken Abelian Higgs
model at finite temperature, the identities acquire a
particularly simple form in momentum space:
$$
\Delta \big( p^2 - m^2 - \Pi(p^2; \rho^2, T) \big) =
  \big( p^2 - m^2 - \Pi(p^2; \rho^2, T) \big) \Delta X \ .
\eqno(3.2)
$$
In the above equation, $m^2=2\mu^2$ is the renormalized
tree-level mass of the Higgs field and a Minkowski metric is used.
The self-energy $\Pi$ can in principle depend on both $p$, the
external four-momentum, and on $\rho$, the VEV of the scalar field.
Clearly, (3.2) implies that the propagator pole-position
as determined by
$$p^2 = m^2 + \Pi(p^2; \rho^2, T)
\eqno(3.3)
$$
is both gauge and parameterization invariant, provided of course
that the corresponding $\Delta X$ is well behaved there.
\par
It must be emphasized that (3.2) is exact.
Thus, the temperature dependent mass as defined by the propagator
pole, and hence the critical temperature, can in
principle be computed in either unitary or renormalizable gauges
to yield the same result.
This is not the case for the curvature of the effective potential,
{\it i.e.,} the inverse-propagator at zero external four-momentum,
which is gauge dependent, in general.
An exception to this arises when the self-energy is independent
of momentum, as happens in the renormalizable gauges to
one-loop-order.
However, when the self-energy does depend on momentum,
it is necessary to use the pole-position definition of the
temperature dependent mass in order to obtain the correct
result for the critical temperature.
\par
There is a line of argument that one might choose to follow in
the hope of expediting the analysis which, unfortunately, leads
to erroneous conclusions.
At the precise value of the critical temperature, the physical
situation demands that the VEV vanish and that the
propagator pole occur at $p^2 = 0$.
Thus, one is tempted to specialize $(3.3)$, and solve
$$
0 = m^2 + \Pi(0;0,T_c) \ ,
\eqno(3.4)
$$
for the critical temperature.
If $(3.4)$ were the case, then it would always be possible to
obtain the correct, gauge invariant, critical temperature from the
effective potential.
However, this argument presumes an exact
knowledge of the self-energy, whereas in practice it is
necessary to evaluate it perturbatively.
We shall now show that the critical temperature
obtained from the pole-position definition of the mass is
gauge and parameterization invariant order-by-order
in any self-consistent expansion (including the loop-expansion).
This is not, however, true for calculations based on (3.4).
\par
There are two distinct and yet equivalent ways in which the
dependence on the VEV can be consistently accounted for in the
self-energy.
The first is to treat the VEV as a parameter to be
solved for perturbatively from the condition that the effective
potential attain its minimum there.
In this case, the self-energy is computed from the
usual set of 1PI diagrams.
The second approach is to include
``tadpole'' diagrams in the calculation of the self-energy.
In this case
one demands that the tree-level tadpole vanish, thus
obtaining the relation given below in $(3.6)$.
The parameter $\rho$ appearing in $(3.3)$ {\it et.seq.,}
is set to its tree-level value to enable consistent
iterative solution for the pole-position.
Higher-order tadpole loop contributions are responsible for
the evolution of the VEV, and their inclusion in
the self-energy takes these effects into account.
In this section, we adopt the former approach because
that is the framework employed by ABV, making it easier for
us to utilize their results at two-loops in the unitary
gauge ({\it c.f.} Eq.(3.6) of Ref.[6]).
However, in the next section, the latter method is chosen, where
only the self-energy
needs to be calculated (with extra tadpole diagrams) rather
than both the self-energy and the effective potential.
\par
One ordinarily employs a perturbative expansion that can be
represented schematically as follows:
$$
\eqalignno{
\Pi &= l \, \Pi_{(1)} + l^2 \, \Pi_{(2)} + \ldots \ ,
&(3.5a)\cr
\rho^2&= \rho^2_{(0)} + l \, \rho^2_{(1)} + l^2 \, \rho^2_{(2)}
+ \ldots \ ,
&(3.5b)\cr
}
$$
Note that the self-energy starts at the first-order in the
expansion, since we assume that the zeroth-order action is
the classical action\footnote{$^1$}{In field
theories that require resummation due to infrared divergences this
is not quite true, but this subtlety has no relevance to the
present discussion.}.
Hence, the lowest-order
scalar VEV must solve the classical field equations.
For the Abelian Higgs model this leads to
$$
\overline{\rho}_{(0)}^2=3m^2/\lambda = 6 \mu^2 / \lambda \ .
\eqno(3.6)
$$
We also assume that renormalization (to absorb
all divergences into the parameters of the theory) has been
carried out consistently.
In particular, the temperature dependent terms in the
perturbative expansions considered here are all finite.
\par
In $(3.5a,b)$, $l$ may be any parameter.
It may be a combination of coupling constants, or a formal
loop-counting parameter, such as $\hbar$.
It is vital that all of the contributions to any given order
in the expansion parameter be properly taken into account.
If this is done, then the pole-position, as
determined by $(3.3)$ is gauge fixing and parameterization
independent {\it order-by-order in $l$}.
Equation $(3.3)$ must be solved with care.
The only self-consistent solution takes the form of a Taylor
expansion in the arbitrary parameter $l$:
$$
\eqalignno{
\pbar^2 &= \pbar^2_{(0)}+l \, \pbar^2_{(1)}
+l^2 \, \pbar^2_{(2)}+\ldots \ ,
&(3.7a) \cr
\overline{\rho}^2 &= \overline{\rho}_{(0)}^2
+ l \, \overline{\rho}_{(1)}^2 + l^2 \, \overline{\rho}_{(2)}^2
+ \ldots \ ,
&(3.7b) \cr
}
$$
where
$$\eqalignno{
\pbar^2_{(0)} &= m^2 \ , &(3.8a)\cr
\pbar^2_{(1)} &= \Pi_{(1)}(p^2; \rho^2 , T)\onpertshell
\ , &(3.8b)\cr
\pbar^2_{(2)} &= \Pi_{(2)}(p^2; \rho^2 , T)\onpertshell
+ \pbar^2_{(1)}
{\partial \Pi_{(1)}\over \partial p^2} \onpertshell
+ \overline{\rho}^2_{(1)}
{\partial \Pi_{(1)}\over\partial \rho^2} \onpertshell
\ ,&(3.8c)\cr
}$$
and so on.
The successive terms in the expansion of $(3.7b)$ are
determined by analysis of the effective potential.
The identities $(3.2)$ guarantee algebraically that
$\pbar^2_{(1)}$, $\pbar^2_{(2)}$, {\it etc.,} are each
separately gauge and parameterization independent.
Note that the correct evaluation of $\pbar^2_{(1)}$
has required setting the self-energy on its bare
mass--shell $p^2=m^2$, as opposed to $p^2=0$.
In renormalizable gauges, since the self-energy is
independent of the external momentum to this order,
this distinction is moot.
In unitary gauge, the distinction is crucial, and accounts for the
discrepancy between the correct value for the critical temperature
and that obtained from an expansion based on $(3.4)$.
Indeed, the gauge dependence identity for the two-point function,
expanded to one-loop (first-order in the perturbative parameter)
yields
$$
\Delta\Pi_{(1)}(0,\rho_{(0)}^2 , T) = - m^2 \Delta X_{(1)} \ ,
\eqno(3.9)
$$
which is not equal to zero in general.
However, $\Delta X_{(1)} = 0$ for renormalizable gauges and
thus the method based on analysis of the effective potential
works at one-loop in this class of gauges.
In unitary gauge the self-energy is momentum dependent, and hence
gauge and parameterization dependent as well, except on-shell.
For cases of this type, the pole-position and associated
self-consistent expansion outlined above must be used in order
to get an invariant estimate for the particle mass and hence the
critical temperature.
\par
In the Abelian Higgs model, it was first shown by Ueda [4] that,
in the high temperature limit,
$$
\Pi_{(1)} = -2 (\lambda/2+3e^2) {T^2\over 12}
- {p^2\over \rhonaught^2} {T^2\over 12}
\eqno{(3.10a)}
$$
in the unitary gauge.
Thus the one-loop correction to the pole-position as obtained
from $(3.8b)$,
$$
\pbar^2_{(1)} =-2\left({\lambda\over 2} +
{\lambda \over 6} + 3 e^2\right) {T^2 \over 12 } \ ,
\eqno(3.10b)
$$
coincides precisely with the value obtained in renormalizable
gauges.
Fortunately, the elegant unitary gauge calculations of ABV
are available to verify that the identities work at two-loops
as well.
The leading-order, high temperature contributions to
the two-loop self-energy sum to
$$
\Pi_{(2)} = {{\lambda}\over{2 \rhonaught^2}}
\left({ T^2\over 12} \right)^2
- {5\over2} {p^2\over \rho_0^4} \left({ T^2\over 12}\right)^2
+ O( T^2 ) \ ,
\eqno{(3.11a)}
$$
while the shift in the scalar VEV is determined to be
$$
\overline{\rho}^2_{(1)} =
- {6\over \lambda} (\lambda/2 + 3e^2) {T^2\over 12}
\ ,
\eqno{(3.11b)}
$$
and thus it is straightforward to verify that
$$
\pbar^2_{(2)} = O( T^2 ) \ .
\eqno(3.12)
$$
The terms of order $T^2$ which contribute to the
second-order correction in $(3.12)$, appear with higher
powers of the coupling constants,
{\it viz.,} $\{ \lambda^2, e^4, \lambda e^2 \}$,
and so they are of higher-order than are the
terms contributing to $\pbar_{(1)}^2$ in $(3.10b)$ above.
\par
Note that throughout this section we treat $l$ as a loop-counting
parameter.
As pointed out by ABV, in unitary gauge, the loop-expansion
corresponds to an expansion in $T^2$.
In principle, therefore, the two-loop contribution could generate
a large correction $O( T^4 )$ to the one-loop result $O( T^2 )$.
The fact that this does not occur is a direct consequence of the
gauge dependence identities, which guarantee that the same
result to any order must be obtained in unitary gauge as in
renormalizable gauges.
In renormalizable gauges it is manifestly clear that the net
$O( T^4 )$ contribution must vanish at two-loop-order.
Note that the expression $(3.8c)$ for $\pbar^2_{(2)}$ involves
both $\Pi_{(2)}$ and derivatives of $\Pi_{(1)}$.
ABV made this observation and interpreted it as a cancellation
between one-loop and two-loop terms.
They were led to conclude that generic calculations in unitary
gauge would require contributions from all higher-loops in order
to agree with results obtained in renormalizable gauges.
In light of the gauge dependence identities, their analysis shows
rather that one must keep all relevant contributions to the
required order in the perturbative (in this case, the loop-counting)
parameter.
\medskip
\goodbreak
\line{\bf 4.  THE COMPLEX $\Phi^4$ MODEL \hfil }
\par
In order to see more clearly the issue of parameterization
dependence in the perturbative computation of the
critical temperature, we consider an interacting complex scalar field
suffering spontaneous symmetry breaking
whose global $U(1)$ symmetry is not gauged.
The Lagrange density is
$$
{{\Lag}} =
( \partial_\mu \Phi )^\dagger ( \partial^\mu \Phi )
+ \mu^2 \vert \Phi \vert^2
- { {\lambda}\over{3!} } ( \vert \Phi \vert^2 )^2  \ .
\eqno{(4.1)}
$$
We introduce a one-parameter family of parameterizations
in terms of two real fields $\{ \phione , \phitwo \}$ and a real
parameter $\epsilon$:
$$
\eqalignno{
\Phi = {1\over{\sqrt{2}}} \big[ \phiR &+ i \phiI \big]  \ ,
&{(4.2)}
\cr
\phiR &= ( 1 - \epsilon ) \, \phione
+ \epsilon \, \phione \cos ( \phitwo / \rhonaught )  \ ,
&{(4.3)}
\cr
\phiI &= ( 1 - \epsilon ) \, \phitwo
+ \epsilon \, \phione \sin ( \phitwo / \rhonaught )  \ .
&{(4.4)}
\cr
}
$$
Inspection of $(4.2-4)$ reveals that the Cartesian
parameterization of Section 2.1 is obtained for $\epsilon = 0$,
while the polar parameterization corresponding to the unitary gauge
in Section 2.2 results when $\epsilon = 1$.
For $0 < \epsilon < 1$, an ``interpolating'' parameterization
ensues.
Both fields $\{ \phione , \phitwo \}$ are chosen to have the same
canonical dimension as $\Phi$; the scale associated with the
angular field is identified here with the zero temperature VEV.
This restriction shall be relaxed in Section 4.1.
\par
The Lagrangian $(4.1)$ may be expressed in terms of
$\{ \phione , \phitwo \}$ by substitution of $(4.2-4)$.
The zero temperature VEV of the scalar field
may be chosen to be concentrated entirely in the $\phione$ mode,
$$
\phione = \rhonaught + \fone \ , \quad
\phitwo = 0 + \ftwo \ .
\eqno{(4.5)}
$$
Substitution of $(4.5)$ into the $\{ \phione , \phitwo
\}$--Lagrangian
corresponds to the usual shift of the fields by their VEVs.
The presence of the trigonometric functions in
$(4.3,4)$ leads to the appearance of vertices to all orders in
$\{ \fone , \ftwo \}$.
However, only those with up to four legs
are required to construct the self-energies to one-loop-order.
The Lagrangian is
$$
\Lag = \Lagcl + \Lagqd + \Lagcu + \Lagqr + \Lagho \ ,
\eqno{(4.6)}
$$
where
$$
\eqalignno{
\Lagcl &=
{1\over2} \mu^2 \rhonaught^2
- {{\lambda}\over{4!}} \rhonaught^4
- \mtwosq \rhonaught \fone
\ ,
\cr
\Lagqd &=
- {1\over2} \fone \big[
              \partial_\mu \partial^\mu + \monesq \big] \fone
- {1\over2} \ftwo \big[
              \partial_\mu \partial^\mu + (1-\epsilon) \mtwosq
\big] \ftwo
\ ,
\cr
\Lagcu &=
- {1\over{3!}} \lambda \rhonaught \fone^3
- {1\over{2}} {{\lambda \rhonaught}\over{3}}
      ( 1 - \epsilon ) \fone \ftwo^2
+ {1\over{2}} {2\over\rhonaught} \epsilon
      \fone ( \partial_\mu \ftwo ) ( \partial^\mu \ftwo )
\ ,
\cr
\Lagqr &=
- {1\over{4!}} \lambda \fone^4
- {1\over{4}} \lambda \bigg[
       {1\over3} ( 1 - \epsilon )
       + \bigg( {{2 \mu^2}\over{\lambda \rhonaught^2}} - {1\over3}
\bigg)
              \epsilon ( 1 - \epsilon )
                           \bigg] \fone^2 \ftwo^2
&(4.7)
\cr
& \qquad \qquad \quad
+ {1\over{4}} \lambda {{2}\over{\lambda \rhonaught^2}} \epsilon^2
              \fone^2 ( \partial_\mu \ftwo ) (\partial^\mu \ftwo )
\cr
& \qquad \qquad \quad
- {1\over{4}} \lambda {{2}\over{\lambda \rhonaught^2}}
 \epsilon ( 1 - \epsilon ) ( \partial_\mu \fone )
          (\partial^\mu \fone ) \ftwo^2
\cr
& \qquad \qquad \quad
- {1\over{4}} \lambda {{4}\over{\lambda \rhonaught^2}}
 \epsilon (1-\epsilon) \fone ( \partial_\mu \fone )
                       \ftwo (\partial^\mu \ftwo )
\cr
& \qquad \qquad \qquad \qquad
- {1\over{4!}} \lambda \bigg[
        ( 1 - \epsilon )^2
       + \bigg( {{3 \mu^2}\over{\lambda \rhonaught^2}} - {1\over2}
\bigg)
              \epsilon ( 1 - \epsilon )
                           \bigg] \ftwo^4
\cr
& \qquad \qquad \qquad \qquad
- {1\over{4}} \lambda {{2}\over{\lambda \rhonaught^2}}
 \epsilon ( 1 - \epsilon ) \ftwo^2 ( \partial_\mu \ftwo )
(\partial^\mu \ftwo )
\ ,
\cr
}
$$
and
$$
\monesq = - \mu^2 + { {\lambda}\over{2} } \rhonaught^2  \ ,
\quad
\mtwosq = - \mu^2 + { {\lambda}\over{6} } \rhonaught^2  \ .
\eqno{(4.8)}
$$
{\baselineskip = 12pt
\midinsert{
%
\vbox{%
\hbox{Propagators:  \hfill}
$$
\matrix{
\Higgsprop  & \hfill  \oneDelta{k}  \hfill =
& \hfill  {{1}\over{k^2 + \monesq}} \hfill
\cr
\Goldprop & \hfill \twoDelta{k} \hfill  =
& \hfill  {{1}\over{k^2}} \hfill
\cr
\Ghostprop & \hfill  1  \hfill  & { }
\cr
}
$$
\hbox{Cubic Vertices:  \hfill}
$$
\matrix{
\vcenter{ \gone } & - g_1 ( q_1 , q_2 , q_3 ) \delta ( \sum q )
\hfill
& g_1 = \lambda \rhonaught
\hfill  \cr
\vcenter{ \gonetwo }& - g_{12} ( q_1 ; q_2 , q_3 ) \delta ( \sum
q ) \hfill
& g_{12} = {{\lambda \rhonaught}\over3} ( 1 - \epsilon )
           - {2\over{\rhonaught}} \epsilon q_2 \cdot q_3
\hfill  \cr
\vcenter{ \gonecbarc }
& - g_{1 \cbar c} ( q_1 ; q_2 , q_3 ) \delta ( \sum q ) \hfill
& g_{1 \cbar c} = {1\over\rhonaught} \epsilon
\hfill  \cr
}
$$
\hbox{Quartic Vertices:  \hfill}
$$
\matrix{
\vcenter{ \lambdaone }
& - \lambda_{1} ( q_1 , q_2 , q_3 , q_4 ) \delta ( \sum q )
\hfill
& \lambda_{1} = \lambda
\hfill  \cr
\vcenter{ \lambdaonetwo }
& - \lambda_{12} ( q_1 , q_2 ; q_3 , q_4 ) \delta ( \sum q )
\hfill
& \lambda_{12} = \lambda
\left[
{
\matrix{
 \phantom{+} {1\over3} ( 1 - \epsilon )
+ {{2}\over{\lambda \rhonaught^2}} \epsilon ( 1 - \epsilon )
q_1 \cdot q_2
     \hfill \cr
+ {{1}\over{\lambda \rhonaught^2}} \epsilon ( 1 - \epsilon )
\left[
       \matrix{ \phantom{+} q_1 \cdot q_3 + q_1 \cdot q_4  \cr
               + q_2 \cdot q_3 + q_2 \cdot q_4  \cr}
\right]
     \hfill \cr
- {{2}\over{\lambda \rhonaught^2}} \epsilon^2
q_3 \cdot q_4
     \hfill \cr
}
}
\right]
\hfill  \cr
\vcenter{ \lambdatwo }
& - \lambda_{2} ( q_1 , q_2 , q_3 , q_4 ) \delta ( \sum q )
\hfill
& \lambda_{2} = \lambda
\left[
{
\matrix{
 \phantom{+} ( 1 - \epsilon )^2
\hfill \cr
+ {{2}\over{\lambda \rhonaught^2}} \epsilon ( 1 - \epsilon )
\left[
{
    \matrix{ \phantom{+} q_1 \cdot q_2 + q_1 \cdot q_3 \cr
            + q_1 \cdot q_4 + q_2 \cdot q_3  \cr
            + q_2 \cdot q_4 + q_3 \cdot q_4  \cr}
}
\right]
     \hfill \cr
}
}
\right]
\hfill  \cr
\vcenter{ \lambdatwocbarc }
& - \lambda_{2 \cbar c} ( q_1 , q_2 ; q_3 , q_4 ) \delta ( \sum q
)
\hfill
& \lambda_{2 \cbar c} = - {{2}\over{\rhonaught^2}}
\epsilon ( 1 - \epsilon )
\hfill  \cr
}
$$
}
{\par\noindent
Table 1:  Feynman rules for the interacting complex scalar model
expressed in terms of the interpolating parameterization.
The solid, dashed and dotted lines denote Higgs (type-one),
Goldstone (type-two) and ghost field propagators, respectively.
The Goldstone boson is massless as a consequence of the tree--level
stability condition.
For the vertices, momenta are positive incoming and are enumerated
in an anticlockwise fashion beginning from the  upper left corner.
Each of the vertices contains an
energy(frequency)-momentum conserving delta-function.
}}
\endinsert
\baselineskip = 15pt }
\par
The presence of the derivative interactions obscures the manifest
renormalizability of the model.
In addition, the field transformations from
$\{ \Phi , \Phi^{*} \} \to \{ \fone , \ftwo \}$
produce a non-trivial Jacobian term in the functional measure,
which can be represented through Grassman-valued ghost fields as
$$
\Laggh = - \cbar \bigg(
1 + {1\over{\rhonaught}} \epsilon \fone
- {1\over{\rhonaught^2}} \epsilon ( 1 - \epsilon ) \ftwo^2 + \ldots
                                           \bigg) c
\ .
\eqno{(4.9)}
$$
The terms which have been omitted (represented by ellipses) are of
higher-order in the fields and hence will not contribute to the
one-loop self-energies.
\par
Feynman rules for this parameterization at finite temperature in the
imaginary time formalism are collected in Table~1.
Appropriate to this formalism, in this section we work with
Euclidean four-momentum $P_\mu$;
the Minkowski mass-shell relation
$p^2=m^2$ of the previous section translates here into
$P^2=-m^2$.
As discussed in Section~3, the viewpoint which shall be
adopted here is to incorporate the effects of the
evolution of the VEV by inclusion of tadpole graphs in the
computation of the self-energies.
By adhering to this prescription, only the self-energy needs
to be computed.
The parametric value of the VEV in $(3.3)$ is set by
the classical Lagrangian, see $(4.10)$, in order that the
perturbative series be well-formed.
Discarding the linear term in the Lagrangian $(4.6)$
corresponds to dispensing with the tree-level tadpole
diagram, and leads to the tree-level stability condition
{\it c.f.,} $(3.6)$:
$$
\mtwosq = 0 \quad \Longleftrightarrow \quad
\monesq = {\lambda\over3} \rhonaught^2 = 2 \mu^2
\quad \Longleftrightarrow \quad
\rhonaught^2 = {{6 \mu^2}\over{\lambda}}
\ .
\eqno{(4.10)}
$$
\par
The collection of diagrams contributing to the self-energy of the
Higgs mode ($\fone$) is assembled in Figure~1, with
the corresponding expressions found in Table~2.
{\baselineskip = 12pt
\midinsert{
\figurethree
{\par\noindent
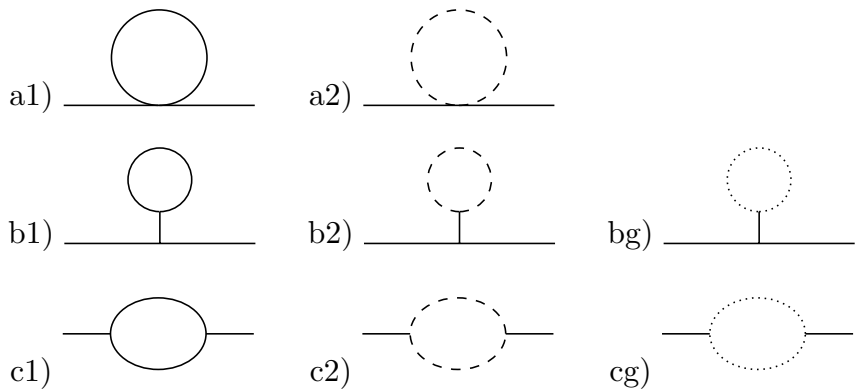
Figure 1:  The self-energy of the Higgs field to one-loop-order
as expressed in terms of the interpolating parameterization. }}
\endinsert
\baselineskip = 15pt }
{\baselineskip = 12pt
\midinsert{
%
%
$$
\matrix{
{\hbox{\underbar{Diagram}}} & - \Pi^{(11)} (P)
&{\hbox{\underbar{Naive Order}}}
\cr
{} & {} & {}  \cr
{\rm a1} & \int dk \, - {\lambda\over2}
\oneDelta{k}  \hfill
&T^2  \hfill  \cr
{\rm a2} & \int dk \,
\left\{
{- {1\over{\rhonaught^2}} \epsilon^2
- {\lambda\over2}
\left(
\matrix{
{1\over3} ( 1 - \epsilon )
- {{2 P^2}\over{\lambda \rhonaught^2}} \epsilon ( 1 - \epsilon )
\hfill  \cr  }
\right)
\twoDelta{k}  }
\right\}  \hfill
&T^4 + T^2 \hfill  \cr
{\rm b1} & \int dk \, {\lambda\over2}
{{\lambda \rhonaught^2}\over{\monesq}}
\oneDelta{k}  \hfill
&T^2  \hfill \cr
{\rm b2} & \int dk \,
\left\{
{ {{\lambda}\over{\monesq}} \epsilon
+ {\lambda\over2}
\left( {{\lambda \rhonaught^2}\over{3 \monesq}} ( 1 - \epsilon )
\right)
\twoDelta{k}  }
\right\} \hfill
&T^4 + T^2  \hfill  \cr
{\rm bg} & \int dk \, - {\lambda\over{\monesq}} \epsilon  \hfill
&T^4  \hfill  \cr
{\rm c1} & \int dk \, {\lambda\over2} ( \lambda \rhonaught^2 )
\oneDelta{k} \oneDelta{P-k}  \hfill
\hfill
&{\hbox{``}} T^0 {\hbox{''}}  \hfill  \cr
{\rm c2} & \int dk \,
\left\{ {
\matrix{
{2\over{\rhonaught^2}} \epsilon^2
& -  {\lambda\over2}
\left[
\matrix{
\left( {{4 P^2}\over{\lambda \rhonaught^2}} \epsilon^2
- {4\over3} \epsilon ( 1 - \epsilon ) \right)
\twoDelta{P - k}  \hfill \cr
- {4\over{\lambda \rhonaught^2}} \epsilon^2 ( P \cdot k )^2
\twoDelta{k} \twoDelta{P - k}  \hfill \cr  }
\right]
\cr
{ } & + \hbox{\ lower order terms}  \hfill  \cr }
} \right\}  \hfill
&T^4 + T^2   \hfill  \cr
{\rm cg} & \int dk \, - {1\over{\rhonaught^2}} \epsilon^2  \hfill
&T^4  \hfill  \cr
}
$$
\medskip
{\noindent
Table~2:  The self-energy contributions to the
$(11)$ two-point function at one-loop order which arise in the
interacting complex scalar model as expressed in the interpolating
parameterization. Here, $\int dk \equiv \sumintk$.
}}
\endinsert
\baselineskip 15 pt }
\par
The effective mass--squared of the Goldstone boson shall also be
explicitly calculated so as to provide a consistency check
on the application of the method and its results.
The set of self-energy diagrams for the Goldstone mode ($\ftwo$)
is found in Figure~2, and their expressions are listed in
Table~3.
{\baselineskip = 12pt
\midinsert{
\figurefour
{\par\noindent
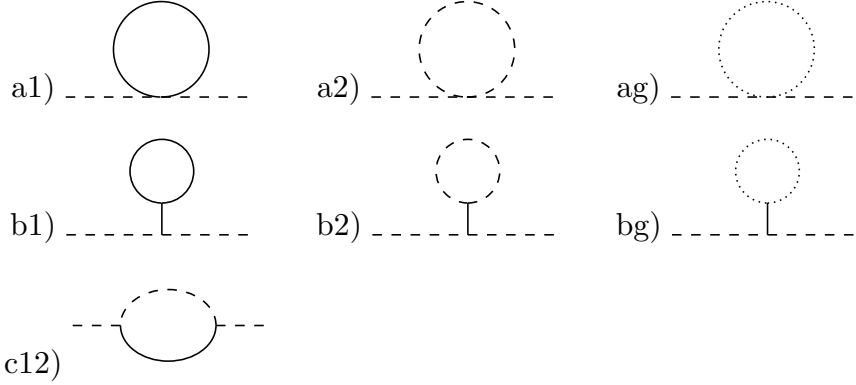
Figure 2.  The self-energy to one-loop-order of the Goldstone
mode, as expressed in terms of the interpolating parameterization.}}
\endinsert
\baselineskip = 15pt }
{\baselineskip = 12pt
\midinsert{
%
$$
\matrix{
{\hbox{\underbar{Diagram}}} & - \Pi^{(22)} (P)
&{\hbox{\underbar{Naive Order}}}
\cr
{} & {} & {}  \cr
{\rm a1} & \int dk \,
\left\{ {
{{1}\over{\rhonaught^2}} \epsilon ( 1 - \epsilon)
-  {\lambda\over2} \left[
\matrix{
{1\over3} ( 1 - \epsilon )
+ {{2 P^2}\over{\lambda \rhonaught^2}} \epsilon^2
\hfill \cr
+ {{2 \monesq}\over{\lambda \rhonaught^2}} \epsilon ( 1 - \epsilon
)
\hfill \cr  }
\right]  \oneDelta{k}
} \right\}  \hfill
&T^4 + T^2   \hfill  \cr
{\rm a2} & \int dk \,
\left\{ {
{{1}\over{\rhonaught^2}} \epsilon ( 1 - \epsilon)
-  {\lambda\over2} \left[
\matrix{
\phantom{+} ( 1 - \epsilon )^{2}
\hfill \cr
- {{2 P^2}\over{\lambda \rhonaught^2}}
             \epsilon ( 1 - \epsilon )
\hfill \cr  }
\right]  \twoDelta{k}
} \right\}  \hfill
&T^4 + T^2   \hfill  \cr
{\rm ag} & \int dk \,
- {2\over{\rhonaught^2}} \epsilon ( 1 - \epsilon )  \hfill
&T^4 \hfill  \cr
{\rm b1} & \int dk \, {\lambda\over2}
\bigg( {{\lambda \rhonaught^2}\over{3 \monesq}} ( 1 - \epsilon )
   + {{2 P^2}\over{\monesq}} \epsilon
\bigg)
\oneDelta{k} \hfill
&T^2  \hfill  \cr
{\rm b2} &
\int dk\, \left\{ {
\matrix{
\matrix{
{{\lambda}\over{3 \monesq}} \epsilon ( 1 - \epsilon )  \hfill \cr
   + {{2 P^2}\over{\rhonaught^2 \monesq}} \epsilon^2   \hfill \cr
}
&
- {\lambda\over2} \left[
\matrix{
\phantom{+}
{{\lambda \rhonaught^2}\over{9 \monesq}} ( 1 - \epsilon )^{2}
\hfill \cr
   - {{2 P^2}\over{3 \monesq}} \epsilon ( 1 - \epsilon )
\hfill \cr }
\right]  \twoDelta{k}
\cr } }  \right\}   \hfill
&T^4 + T^2  \hfill  \cr
{\rm bg} & \int dk \,
- \bigg( {{\lambda}\over{3 \monesq}} \epsilon ( 1 - \epsilon )
   + {{2 P^2}\over{\rhonaught^2 \monesq}} \epsilon^2
\bigg)   \hfill
&T^4 \hfill  \cr
{\rm c12} & \int dk \, \lambda
{4\over{\lambda \rhonaught^2}} \epsilon^2 ( P \cdot k )^2
\oneDelta{k} \twoDelta{P - k}  \hfill
&T^2  \hfill \cr
}
$$
\medskip
\noindent
Table 3:  The self-energy contributions to the $(22)$ two-point
function at one-loop-order which arise in the interacting complex
scalar model as expressed in the interpolating parameterization.
}
\endinsert
\baselineskip = 15pt }
\par
In Tables~2 and 3, the leading naive degrees of temperature
dependence associated with each diagram are given.
We note that the
vacuum contributions which are quartically divergent (associated
with the $O(T^4)$ pieces) all cancel.
\par
To evaluate each of these self-energies, the field
is chosen to be at rest with respect to the heat bath.
In the high temperature limit, the following results hold:
$$
\eqalignno{
\int dk \, {1\over{k^2 + m^2}} \atthermal &=
{{J ( \beta m )}\over{\beta^2}}
\ ,
\cr
\int dk \, {{( P \cdot k )^2}\over{ \big( k^2 + m^2 \big)
                                   \big( (P-k)^2 + m^2 \big) }}
 \atthermalatrest &=
{1\over2} P^2 {{J ( \beta m )}\over{\beta^2}}
+ O ( \beta^{-1} )
\ ,
&{(4.11)}
\cr
\int dk \, {{( P \cdot k )^2}\over{ \big( k^2 + m^2 \big)
                                   \big( (P-k)^2 + M^2 \big) }}
\atthermalatrest &\simeq
{1\over2} P^2 {{J ( \beta m )}\over{\beta^2}}
+ O ( \beta^{-1} )
\ ,
\cr
}
$$
where
$$
J ( c ) = {1\over{2 \pi^2 }} \int_{0}^{\infty} dx
{{x^2}\over{\sqrt{x^2 + c^2}}}
{{1}\over{ e^{\sqrt{ x^2 + c^2 }} - 1}}
\ .
\eqno{(4.12)}
$$
Moreover, in the high temperature limit $\beta m \rightarrow 0$,
$$
J ( \beta m ) = {1\over{12}} + O ( \beta m )
\ .
\eqno{(4.13)}
$$
This limit, as well as $\beta M\rightarrow 0$,
is already assumed in equation $(4.11)$.
\par
The order-parameter for the symmetry restoring phase transition
is the effective mass--squared of the Higgs field,
$$
\moneeffsq = \monesq + \Pitottherm{11} \atresthightemponshell
\ .
\eqno{(4.14)}
$$
Note that renormalization is assumed to have taken place, and
$\monesq$ represents the physical value of the zero temperature
Higgs boson mass--squared.
In the high temperature (weak coupling) limit, we find
$$
\moneeffsq = 2 \mu^2 - {{\lambda}\over{9 \beta^2}}
\bigg\{ 1 - {1\over4} \bigg( 1 + {{P^2}\over{\monesq}} \bigg)
           \bigg[\epsilon \big(2 \epsilon -1 \big) \bigg]
      \bigg\}  \onshell
+ \quad O ( \beta^{-1} )
\ .
\eqno{(4.15)}
$$
Thus, on-shell at $P^2 = - \monesq$
the coefficient of the $\epsilon$-dependent term vanishes
identically,
and the final expression for the order-parameter is
$$
\moneeffsq = 2 \mu^2 - {{\lambda}\over{9 \beta^2}}
+ O ( \beta^{-1})
\ .
\eqno{(4.16)}
$$
Equation $(4.16)$ is valid for the entire $\epsilon$-family of
parameterizations and yields the correct critical temperature
({\it c.f.,} $(2.5 - 6)$) in all cases.
This is the central result of our analysis of the interacting
complex scalar model and has provided
a practical illustration of the gauge-dependence identities.

Note that in the $\epsilon = 0$ limit of the general result
$(4.15)$
the term in square brackets vanishes, and the correct physical
one-loop result for the value of the order-parameter $(4.16)$ then
follows independently of the application of the on-shell condition.
This is due to the absence of derivative interactions
in the Cartesian parameterization.
Thus, the method of analysis based upon the effective potential
approach as described in Section~2 works in the Cartesian case
to one-loop-order.
Also note that for the polar case $\epsilon=1$, evaluating the
order parameter off-shell at $P_\mu=0$ results in the (erroneous)
result of the unitary gauge calculated from the effective
potential alone:
$$
\moneeffsq \atpzero
= 2 \mu^2 - {{\lambda}\over{12 \beta^2}} + O ( \beta^{-1} )
\ .
\eqno{(4.17)}
$$
\par
As a test of the validity of the method and the consistency of
the calculations, consider the expression for the effective
mass--squared of the Goldstone mode:
$$
\mtwoeffsq = \mtwosq + \Pitottherm{22} \atresthightemponshell
\ .
\eqno{(4.18)}
$$
We expect that the Goldstone boson is massless
in the broken phase below $T_c$, and indeed collecting all terms
in the high temperature limit yields
$$
\mtwoeffsq = 0 - {{\lambda}\over{12 \beta^2}}
\bigg\{
{{P^2}\over{\lambda \rhonaught^2}} \epsilon ( 5 - \epsilon )
      \bigg\}  \onshell
+ \quad O ( \beta^{-1} )
\ ,
\eqno{(4.19)}
$$
which vanishes at $P^2 = - \mtwosq = 0$,
independent of $\epsilon$. Thus, the Goldstone boson remains
massless at one-loop to the order at which it is calculated.
\par
\smallskip
\goodbreak
\line{\bf 4.1)
The Case of an Arbitrary Scale For the Goldstone Mode \hfil }
\par
The calculation described above was undertaken with a
parameterization which carried an explicit dependence
on the VEV at zero temperature.
In this subsection, a two-parameter family
of parameterizations will be constructed in which the scale
$\rho$ is not identified with the VEV of the field:
$$
\eqalignno{
\Phi = {1\over{\sqrt{2}}} \big[ \phiR &+ i \phiI \big]  \ ,
&{(4.20)}
\cr
\phiR &= ( 1 - \epsilon ) \, \phione
+ \epsilon \, \phione \cos ( \phitwo / \rho )  \ ,
&{(4.21)}
\cr
\phiI &= ( 1 - \epsilon ) \, \phitwo
+ \epsilon \, \phione \sin ( \phitwo / \rho )  \ .
&{(4.22)}
\cr
}
$$
This can be written in terms of the VEV $\rho_0$ by introduction
of a dimensionless parameter:
$$
\rho = \eta \rhonaught \ , \quad 0 < \eta < \infty
\ .
\eqno{(4.23)}
$$
A frequently occurring combination of $\eta$ and $\epsilon$
shall be denoted by $b$, where
$$
b = \eta + \epsilon - \eta \epsilon =
\eta + ( 1 - \eta ) \epsilon = \epsilon + ( 1 - \epsilon ) \eta
\ .
\eqno{(4.24)}
$$
Although not necessary in this context, it is possible
to choose a parameterization for which the standard
normalization of the kinetic term for the Goldstone field results;
this can be accomplished by using
$$
\eqalignno{
\phiR &= ( 1 - \epsilon ) \, \phione
+ \epsilon \, \phione \cos ( \phitwo / \rhonaught b )  \ ,
&{(4.25)}
\cr
\phiI &= {{( 1 - \epsilon ) \eta}\over{b}} \, \phitwo
+ \epsilon \, \phione \sin ( \phitwo / \rhonaught b )  \ ,
&{(4.26)}
\cr
}
$$
rather than the parameterization of (4.21 -- 4.22).
VEVs are then assigned to the fields as
$$
\phione = \rhonaught + \fone \ , \quad
\phitwo = 0 + \ftwo \ .
\eqno{(4.27)}
$$
The Feynman rules for this parameterization are collected in
Table 4.
Fortunately, no new vertices are required beyond those listed in
Table~1, but the vertex factors are more complicated.
As before, the
tree--level stability condition $(4.10)$ arises again when
elimination of the linear term in the Lagrangian is imposed.
{\baselineskip = 12pt
\midinsert{
%
\vbox{%
\hbox{Propagators:  \hfill}
$$
\matrix{
\Higgsprop  & \hfill \oneDelta{k}  \hfill =
& \hfill  {{1}\over{k^2 + \monesq}} \hfill
\cr
\Goldprop & \hfill  \twoDelta{k} \hfill  =
& \hfill  {{1}\over{k^2}} \hfill
\cr
\Ghostprop & \hfill  1  \hfill  & { }
\cr
}
$$
\hbox{Cubic Vertices:  \hfill}
$$
\matrix{
\vcenter{ \gone } & - g_1 ( q_1 , q_2 , q_3 ) \delta ( \sum q )
\hfill
& g_1 = \lambda \rhonaught
\hfill  \cr
\vcenter{ \gonetwo }& - g_{12} ( q_1 ; q_2 , q_3 ) \delta ( \sum
q ) \hfill
& g_{12} = {{\lambda \rhonaught}\over3} {{ b^2 - \epsilon
}\over{b^2}}
           + {{\epsilon ( 1 - b )}\over{\rhonaught b^2}}
                  [ q_1 \cdot ( q_2 + q_3 ) ]
           - {{2 \epsilon}\over{\rhonaught b}} q_2 \cdot q_3
\hfill  \cr
\vcenter{ \gonecbarc }
& - g_{1 \cbar c} ( q_1 ; q_2 , q_3 ) \delta ( \sum q ) \hfill
& g_{1 \cbar c} = {\epsilon\over{\rhonaught b}}
\hfill  \cr
}
$$
\hbox{Quartic Vertices:  \hfill}
$$
\matrix{
\vcenter{ \lambdaone }
& - \lambda_{1} ( q_1 , q_2 , q_3 , q_4 ) \delta ( \sum q )
\hfill
& \lambda_{1} = \lambda
\hfill  \cr
\vcenter{ \lambdaonetwo }
& - \lambda_{12} ( q_1 , q_2 ; q_3 , q_4 ) \delta ( \sum q )
\hfill
& \lambda_{12} = \lambda
\left[
{
\matrix{
 \phantom{+} {1\over3} {{ b^2 - \epsilon }\over{b^2}}
- {{4 \epsilon ( 1 - b )}\over{3 b^2}}
     \hfill \cr
+ {{2 \epsilon ( 1 - \epsilon )}\over{\lambda \rhonaught^2 b^2}}
q_1 \cdot q_2
     \hfill \cr
+ {{\epsilon ( 1 - \epsilon )}\over{\lambda \rhonaught^2 b^2}}
\left[
       \matrix{ \phantom{+} q_1 \cdot q_3 + q_1 \cdot q_4  \cr
               + q_2 \cdot q_3 + q_2 \cdot q_4  \cr}
\right]
     \hfill \cr
- {{2 \epsilon^2}\over{\lambda \rhonaught^2 b^2}}
q_3 \cdot q_4
     \hfill \cr
}
}
\right]
\hfill  \cr
\vcenter{ \lambdatwo }
& - \lambda_{2} ( q_1 , q_2 , q_3 , q_4 ) \delta ( \sum q )
\hfill
& \lambda_{2} = \lambda
\left[
{
\matrix{
 \phantom{+} {{( b^2 - \epsilon )^2}\over{b^4}}
     \hfill \cr
+ {{2 \epsilon ( 1 - \epsilon ) \eta}\over{\lambda \rhonaught^2
b^4}}
\left[
{
    \matrix{ \phantom{+} q_1 \cdot q_2 + q_1 \cdot q_3  \cr
                       + q_1 \cdot q_4 + q_2 \cdot q_3  \cr
                       + q_2 \cdot q_4 + q_3 \cdot q_4 \cr}
}
\right]
     \hfill \cr
}
}
\right]
\hfill  \cr
\vcenter{ \lambdatwocbarc }
& - \lambda_{2 \cbar c} ( q_1 , q_2 ; q_3 , q_4 ) \delta ( \sum q
)
\hfill
& \lambda_{2 \cbar c} =
- {{\epsilon ( 1 - \epsilon ) ( 1 + \eta )}\over{\rhonaught^2 b^3}}
\hfill  \cr
}
$$
}
\par\noindent
Table 4:  Feynman rules for the interacting complex scalar model
expressed in terms of the modified interpolating parameterization.
The notation and conventions are the same as those used in Table~1.}
\endinsert
\baselineskip = 15pt }
\par
The same diagrams listed in Figure~1 contribute to the self-energy
in this more general parameterization.
Evaluation of these diagrams in the high-temperature limit leads
to the following expression for the order-parameter:
$$
\moneeffsq = 2 \mu^2 - {{\lambda}\over{9 \beta^2}}
\bigg\{ 1 - {1\over{4 b^2}} \bigg( 1 + {{P^2}\over{\monesq}} \bigg)
    \bigg[\epsilon \left( {{2 \epsilon}\over{b}} - 1 \right) \bigg]
\bigg\}  \onshell
+ \quad O ( \beta^{-1} )
\ .
\eqno{(4.28)}
$$
There are several things to note about this equation.
The first is that setting the external four-momentum on--shell,
$P^2 = - \monesq$, causes the coefficient of the $\epsilon$-
and $b$-dependent terms to vanish.
Thus, the correct physical result is obtained for all values
of the parameters.
The second feature of $(4.28)$ is that in the limit
in which the scale is chosen to equal the VEV, {\it viz.,}
$\eta = 1$ and hence $b = 1$, then the result reduces to
the one obtained earlier in $(4.15)$.
The third aspect of $(4.28)$ we mention is
that its structure guarantees that the correct physical result
is obtained to this order both on-- and off--shell in the
Cartesian parameterization, $\epsilon = 0$, as expected.
Finally, it is noted that when the polar parameterization,
$\epsilon = 1$, is chosen, then $b=1$, and the result
is manifestly independent of the scale chosen for the
angular Goldstone field.
This fortuituous independence enabled the unitary gauge
puzzle to be more precisely formulated when it was first
discovered.
\par
Finally, the Goldstone boson's thermal self-energy corrections
may be calculated from the diagrams of Figure~2.
The result is:
$$
\mtwoeffsq = 0 - {{\lambda}\over{12 \beta^2}}
\bigg\{
P^2 G( \epsilon, \eta, b )
      \bigg\}  \onshell
+ \quad O ( \beta^{-1} )
\ ,
\eqno{(4.29)}
$$
where $G( \epsilon, \eta, b )$ is a function of its
parameteric arguments.
Thus, as was the case in $(4.19)$, evaluating $(4.29)$ on-shell
at $P^2=0$ shows that the Goldstone boson remains
massless at one-loop-order
independent of the precise details of the parameterization.
 \medskip
\goodbreak
\line{\bf 5.  CONCLUSIONS \hfil }
\par
We have shown that a complete resolution to the so-called unitary
gauge puzzle is obtained by examining the Ward identities that
govern the gauge and parameterization dependence of the quantities
under consideration. These show directly that the discrepancy
between the unitary gauge and renormalizable gauge results for the
critical temperature derive not from any intrinsic pathologies of
the unitary gauge, but from the momentum dependence of the
self-energy. We argued that gauge invariant results for the
critical
temperature can be obtained order-by-order in any self-consistent
perturbative expansion in any gauge, including unitary gauge. This
was verified explicitly at the two-loop level to order $T^4$ in
scalar QED, and at the one-loop level in a family of
parameterizations in complex $\phi^4$ theory. The latter calculation
also stresses the important, but sometimes forgotten, fact that the
unitary gauge puzzle is concerned more with parameterization
invariance than with gauge invariance, although the two issues are
deeply related.
\par
It must be stressed that our analysis does more than re-interpret
the two-loop calculations of ABV. First of all, our
arguments, admittedly formal, provide an all-order proof that the
cancellation observed
by ABV in unitary gauge at the two-loop order occurs to all orders
in perturbation theory. Secondly, our analysis provides criteria
that can be used to determine whether or not a given perturbative
calculation should be expected to yield gauge invariant results.
In particular, it is necessary to start at lowest-order from a gauge
invariant quantity (such as the pole-position of a physical
propagator), and perform a self-consistent expansion.
In the present paper the loop-counting
parameter was used, but of course any
convenient parameter may be used (coupling constant, temperature,
{\it etc.}).
It is, however, important to note that although any such
self-consistent
expansion will guarantee gauge independence order-by-order,
the question of accuracy must be addressed independently.
The weakness of unitary guage is that this question is obscured
due to the presence of non-renormalizable interactions introduced
by the non-linear reparameterization.
The utility of the gauge dependence
identities lies in the fact that they guarantee that the error
estimate can be done in any convenient gauge, such as
renormalizable gauge, while
the actual calculations can be performed in any other gauge,
because the same answer will be obtained {\it order-by-order} in
all gauges. Thus, for example, the accuracy of the
leading one-loop result
for the critical temperature in renormalizable gauges implies that
higher-loop contributions in unitary gauge will also be negligable.
\medskip
\goodbreak
\line{\bf Acknowledgements \hfil}
We thank R. Jackiw, E. Braaten and S. Vokos
for valuable discussions. This work was supported in part by the
Natural Sciences and Engineering Research Council of Canada,
the United States Department of Energy under contract
\#DE-AC02-76ER03069 and cooperative agreement \#DE-FC02-94ER40818,
the National Science Foundation under
Grant No. PHY89-04035, and
the Centre International des Etudiants et Stagiaires de France.

\medskip\medskip
\vfill\eject
\line{\bf REFERENCES \hfil}
\item{1.} D.A. Kirzhnits and A.D. Linde, Phys. Lett. {\bf B42},
471 (1972).
\item{2.} L. Dolan and R. Jackiw, Phys. Rev. {\bf D9}, 3320 (1974).
\item{3.} S. Weinberg, Phys. Rev. {\bf D9}, 3357 (1974).
\item{4.} Y. Ueda, Phys. Rev. {\bf D23}, 1383 (1981).
\item{5.} M. Chaichian, E.J. Ferrer and V. de la Incera, Nucl. Phys.
{\bf B362}, 616 (1991).
\item{6.} P. Arnold, E. Braaten and S. Vokos, Phys. Rev. {\bf D46},
3576 (1992).
\item{7.} L. Dolan and R. Jackiw, Phys. Rev. {\bf D9}, 2904 (1974).
\item{8.} G. Kunstatter and H.P. Leivo, Phys. Lett. {\bf B183}, 75
(1987).
\item{9.} G.A. Vilkovisky, {\it in} Quantum Theory of Gravity, ed.
S.M. Christensen (Adam Hilger, Bristol, 1984); B.S. DeWitt, {\it
in}, Quantum Field Theory and Quantum Statistics, ed. I.A. Batalin,
C.J. Isham and G.A. Vilkovisky, (Adam Hilger, Bristol, 1987).
\item{10} R. Kobes, G. Kunstatter and D.J. Toms, {\it in}, TEV
Physics, ed. G. Domokos and S. Kovesi-Domokos, (World Scientific,
Singapore, 1988).
\item{11.} I. Lawrie, J. Phys. {\bf A} (in press); R. Jackiw and
G. Amelino-Camelia, in {\it Banff/CAP Workshop on Thermal Field Theory}
(World Scientific, Singapore, 1994).
\item{12.} N.K. Nielsen, Nucl. Phys. {\bf B101}, 173 (1975); I.J.R.
Aitchison and C.M. Fraser, Ann. of Phys. {\bf 156}, 1 (1984).
\item{13.} R. Kobes, G. Kunstatter and A. Rebhan, Phys. Rev. Lett.
{\bf 64}, 2992 (1990); Nucl. Phys. {\bf B355}, 1 (1991).
 \bye